\documentclass{article}
\usepackage{spconf,amsmath,graphicx}
\usepackage{amsfonts}
\usepackage{enumitem}
\setlist{nosep, leftmargin=14pt}
\usepackage{hyperref}
\usepackage{mwe} 
\usepackage{multirow}
\usepackage[table,xcdraw]{xcolor}
\usepackage{orcidlink}

\title{Intensity-based 3D motion correction for cardiac MR images}
\name{Nil Stolt-Ansó $^{\star (1,2)}$\orcidlink{0009-0001-4457-0967},  Vasiliki Sideri-Lampretsa $^{\star (2,3)}$ \thanks{$^{\star}$Equal contribution.}\orcidlink{0000-0003-0135-7442}, Maik Dannecker $^{(2,3)}$\orcidlink{0000-0001-9012-9606}, Daniel Rueckert $^{(1,2,3,4)}$\orcidlink{0000-0002-5683-5889}}
\address{\small 1. Munich Center for Machine Learning, Technical University Munich\\
\small 2. School of Computation, Information and Technology, Technical University Munich\\
\small 3. Klinikum Rechts der Isar, Technical University Munich\\
\small 4. Department of Computing, Imperial College London
}
\begin{document}
%
\maketitle
\begin{abstract}
Cardiac magnetic resonance (CMR) image acquisition requires subjects to hold their breath while 2D cine images are acquired. This process assumes that the heart remains in the same position across all slices. However, differences in breathhold positions or patient motion introduce 3D slice misalignments. In this work, we propose an algorithm that simultaneously aligns all SA and LA slices by maximizing the pair-wise intensity agreement between their intersections. Unlike previous works, our approach is formulated as a subject-specific optimization problem and requires no prior knowledge of the underlying anatomy. We quantitatively demonstrate that the proposed method is robust against a large range of rotations and translations by synthetically misaligning \(10\) motion-free datasets and aligning them back using the proposed method.

\end{abstract}
\begin{keywords}
cardiac magnetic resonance imaging, alignment, contour-free, single-subject, motion correction
\end{keywords}
\section{Introduction}
\label{sec:introduction}

Cardiac magnetic resonance (CMR) imaging is typically acquired along acquisition planes aligned with the cardiac anatomy, each intended to allow the extraction of different clinically relevant measurements of target regions. Most notably, the long axis (LA) and short axis (SA) are planes that aim to visualize the left ventricle (LV) to derive longitudinal and radial motion information, respectively. Although the acquisition of slices is typically performed in a fixed scanner coordinate system, each slice is an independent acquisition originating from a separate breathhold. This makes anatomical misalignments between slices a common occurrence due to variations in the heart's location across acquisitions.

\begin{figure}[htb]
  \centering
  \centerline{\includegraphics[trim={0 5cm 0 0},clip, width=\linewidth]{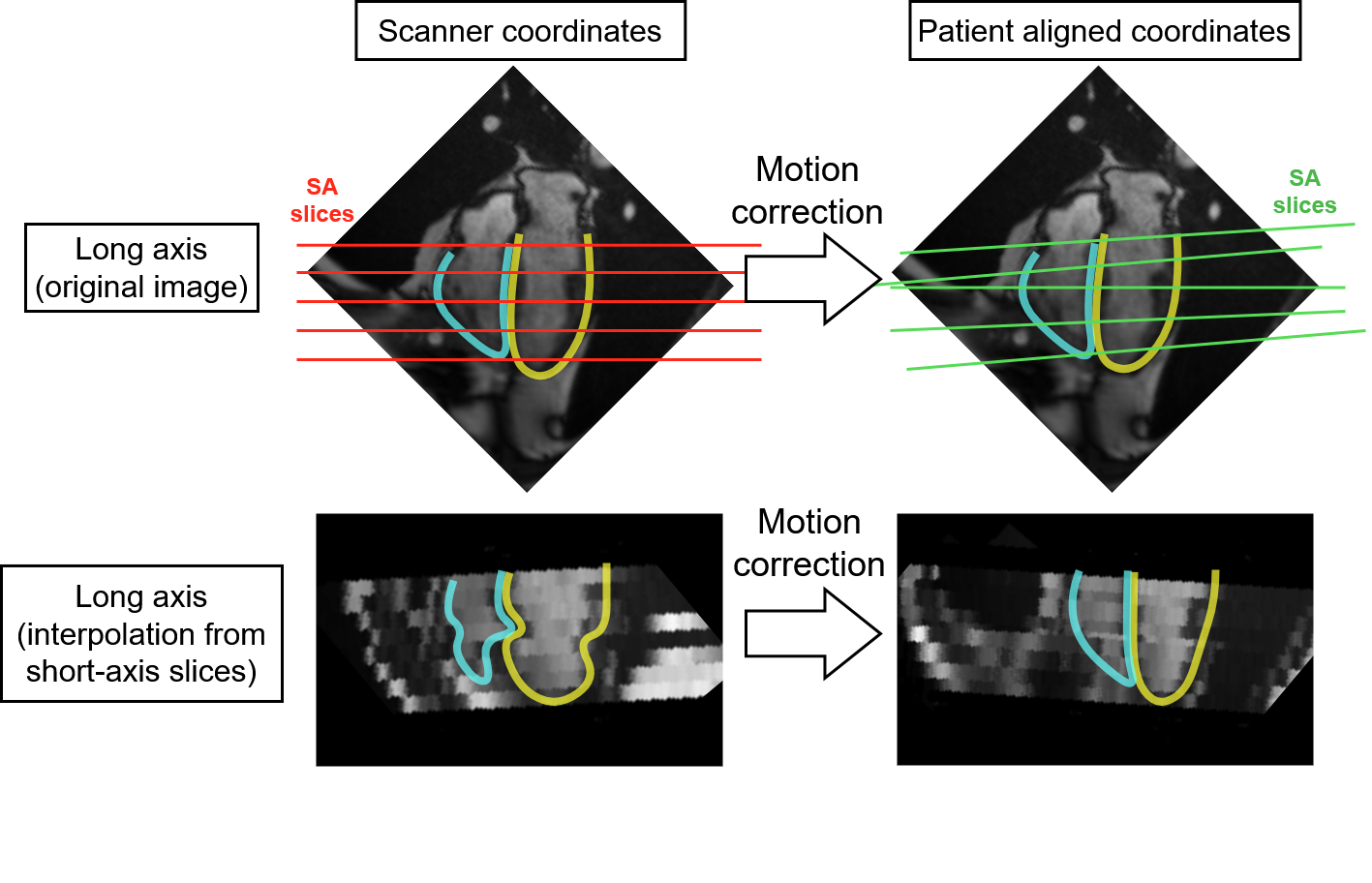}}
\caption{Diagram of resulting optimization on slice alignment. Ventricle contours are hand-drawn for illustrative purposes. \textbf{Top row}: LA 4-chamber image with intersections lines of SA slices with original orientations (left/red) and optimized orientations (right/green). \textbf{Bottom row}: Nearest neighbour interpolation of LA 4-chamber view from SA slices with original orientations (left) and optimized orientations (right).}
\label{fig:alignment}
\end{figure}

\textbf{Acquisition planes.} The long axis is the line that passes from the center of the mitral valve to the apex of the heart. The LV myocardium typically has three long-axis imaging planes: 2-chamber, 3-chamber, and 4-chamber views. These are named after the number of ventricles and atria they intersect and are meant to capture different sections of the myocardium. The short axis is perpendicular to the long axis, visualizing myocardial cross-sections at various ventricular steps. These slices are stacked for volumetric measurements.

\textbf{Motion correction.} Considerable research efforts using image registration have been made to correct these misalignments \cite{Chandler2008CorrectionOM,Tarroni2018ACA,Su2014meshAlign,Villard2016CorrectionOS,Ltjnen2004CorrectionOM,Sinclair2017FullyAS,Yang20173DMM}. In~\cite{Chandler2008CorrectionOM}, Chandler et al. aim to correct misaligned cardiac anatomy in multi-slice SA images by rigidly registering stacks of two slices to a high-resolution 3D MR axial cardiac volume. Although they demonstrate good results in synthetic and real misaligned data, the method requires a high-resolution 3D cardiac volume, which is not always available.

\begin{figure*}[htb]
  \centering
  \centerline{\includegraphics[width=\textwidth]{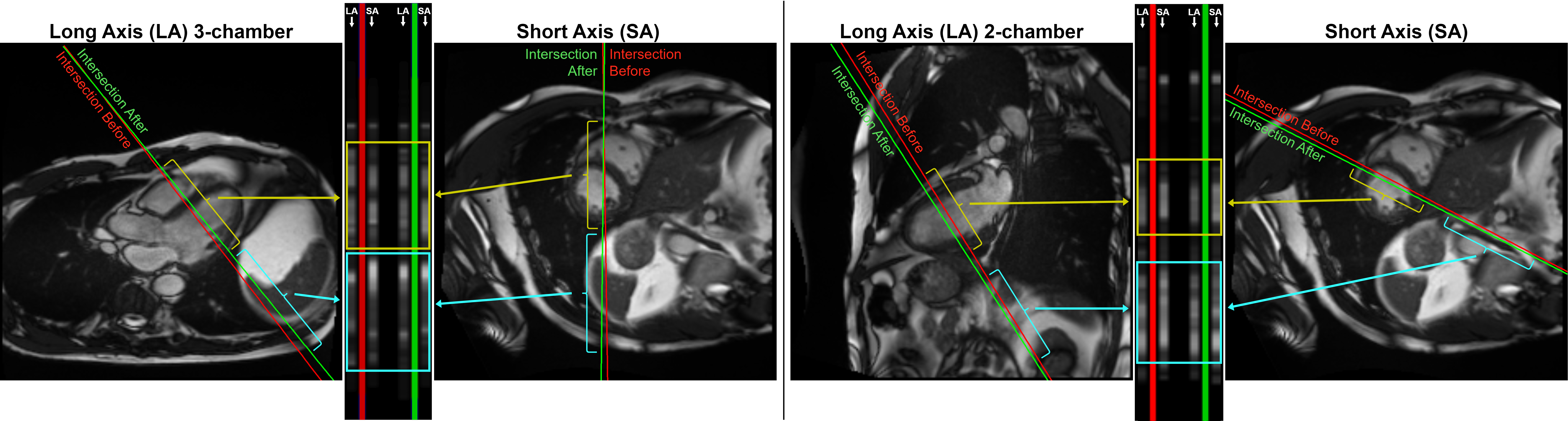}}
\caption{Result of intensity interpolation at slice intersections before (red) and after (green) optimization. Two slice pairs are displayed: short axis (SA) to long axis (LA) 3-chamber (left), and short axis to long axis 2-chamber (right). Red and green lines depict the intersection lines of each image pair. Each pair's interpolated intersection intensities are displayed side-by-side on the vertical block. Blue segments depict padding points outside of image bounds. Areas of significant change are shown in yellow (ventricles) and cyan (stomach).}
\label{fig:sample}
\end{figure*}
Another approach is to utilize shape information~\cite{Su2014meshAlign, Tarroni2018ACA,Sinclair2017FullyAS,Yang20173DMM}. Su et al.~\cite{Su2014meshAlign} combine 3D meshes and apply the constraint that the epicardial surface must remain smooth to generate a time-series model of the heart. The 3D meshes are created from border-delineated MRI data at every time frame of the cardiac cycle. Tarroni et al.~\cite{Tarroni2018ACA} corrects inter-slice respiratory motion in SA CMR image stacks utilizing probabilistic segmentation maps (PSMs) of the left ventricular (LV) cavity generated with hybrid decision forests. PSMs are generated for each slice of the SA stack and rigidly registered in-plane to a target PSM. Both these methods utilize shape information that is not always available or might not generalize outside the intended subject cohort, limiting the methods' applicability. A more recent approach \cite{Villard2016CorrectionOS} proposes a slice-to-slice group-registration framework based on the similarity of the local phase vectors of the images. As registration is notorious for being susceptible to local minima,  they fix all slices and only allow one slice to "move" at a time.

\textbf{Contributions.} In this work, we propose a method to mitigate the effect of inter-slice motion for all SA and LA slices simultaneously by optimizing the 3D rotation and translation parameters on sampled intensities along slice intersections. Unlike other approaches, our approach exclusively utilizes the image intensity information, which requires no prior knowledge about the underlying anatomy. We demonstrate that our GPU-accelerated approach can reliably converge in seconds despite large perturbations on the initial alignment parameters. Our contributions can be summarized as follows:
\begin{itemize}
    \item We propose a solution for motion correction in CMR that jointly aligns all SA and LA slices simultaneously.
    \item We utilize all SA-LA and LA-LA intersections and minimize the intensity differences along the intersection lines without incorporating any anatomical information.
    \item We evaluate our method quantitatively recovering synthetic rigid deformations using aligned CMR scans as a golden standard.
\end{itemize}

\section{Method}
\label{sec:method}

With this method, we try to address SA-LA and LA-LA misalignments due to patient motion in the scanner with respect to their anatomy. To do that, we assume that any deformation applied to a slice during acquisition is strictly rigid. More precisely, our approach corrects for \(3D\) rotation and translation while it does not model scaling or shearing deformations. Our rigid rotation and translation matrices are defined as \(R(\boldsymbol{\theta})\) and \(T(\boldsymbol{t}) \:\), where $\boldsymbol{\theta}$ and $\boldsymbol{t}$ are 3D vectors corresponding to rotation angles and translation coefficients for a given slice.

When two intersecting slices are perfectly aligned, a metric computing similarity of the intensities along the intersection line of the two planes should be at a global maximum. We formulate the alignment problem as a constraint optimization of parameters $\phi\in\mathbb{R}^{N\times6}$ that aims to jointly minimize the pairwise intensity differences at slice intersections (where $N$ is the number of slices and 6 is the total rotation and translation deformation parameters in 3D). We use the summation of pair-wise L2 intensity differences as a similarity metric between the intensities along slice intersections, which minimizes the influence of shifts in intensities across slices.  

The set of slice pairs is based on all possible LA-LA slice combinations and LA-SA slice combinations of a given subject. We do not consider SA-SA combinations since their planes do not overlap significantly. For every slice pair $(A, B)$, we compute their plane equations given their current rotation and translation parameters and thereby obtain the intersection line $d^{AB}$. We define a sampling center $\boldsymbol{c}$ by finding the closest points $\boldsymbol{c^A}$ and $\boldsymbol{c^B}$ along line $d^{AB}$ to the centers of images $A$ and $B$, and subsequently taking the mid-point of the two. Using $\boldsymbol{c}$, we can define a sampling segment $K\in \mathbb{R}^{p\times 3}$ in 3D space, where $p$ is the number of sampling points. We set $p$ to 100 samples (50 points in each direction of $\boldsymbol{c}$ along $d^{AB}$) and the distance between samples to be 5 millimeters in world space.

\begin{figure*}[ht!]
  \centering
  \centerline{\includegraphics[width=\linewidth]{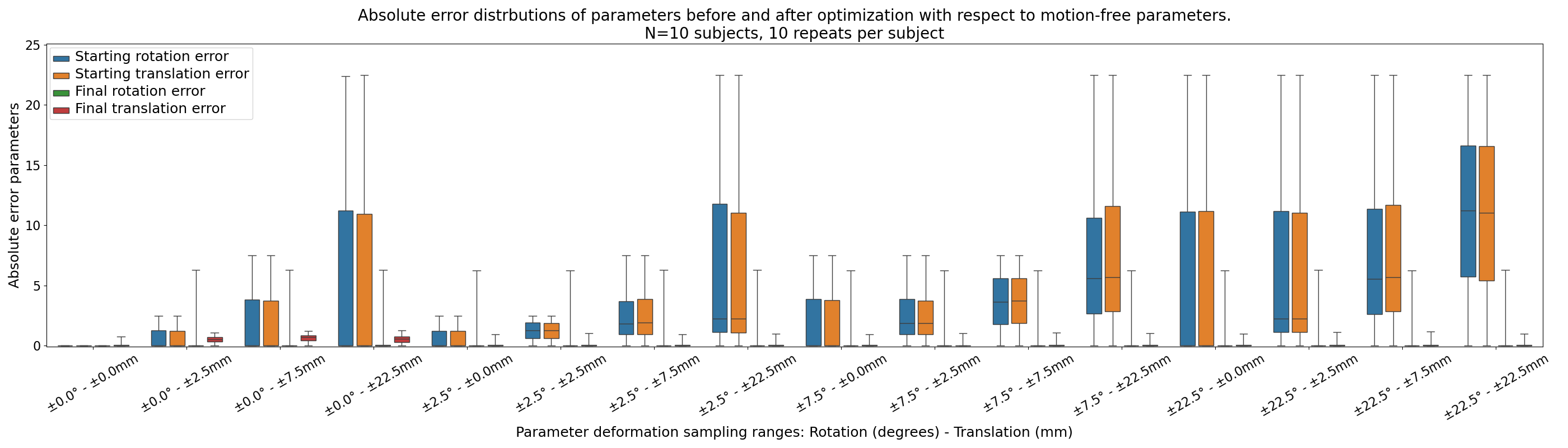}}
  \caption{Distributions of absolute error in parameters for all runs (with respect to motion-free parameters). Absolute errors are presented for parameters before and after optimization for both rotation and translation parameters. Results are shown under various ranges of uniformly sampled initial rigid transformations. Dataset consisted of 10 aligned subjects, each of which was randomly misaligned 10 times per condition.}
\label{fig:results_before_after}
\end{figure*}

\begin{figure*}[ht!]
  \centering
  \centerline{\includegraphics[width=\linewidth]{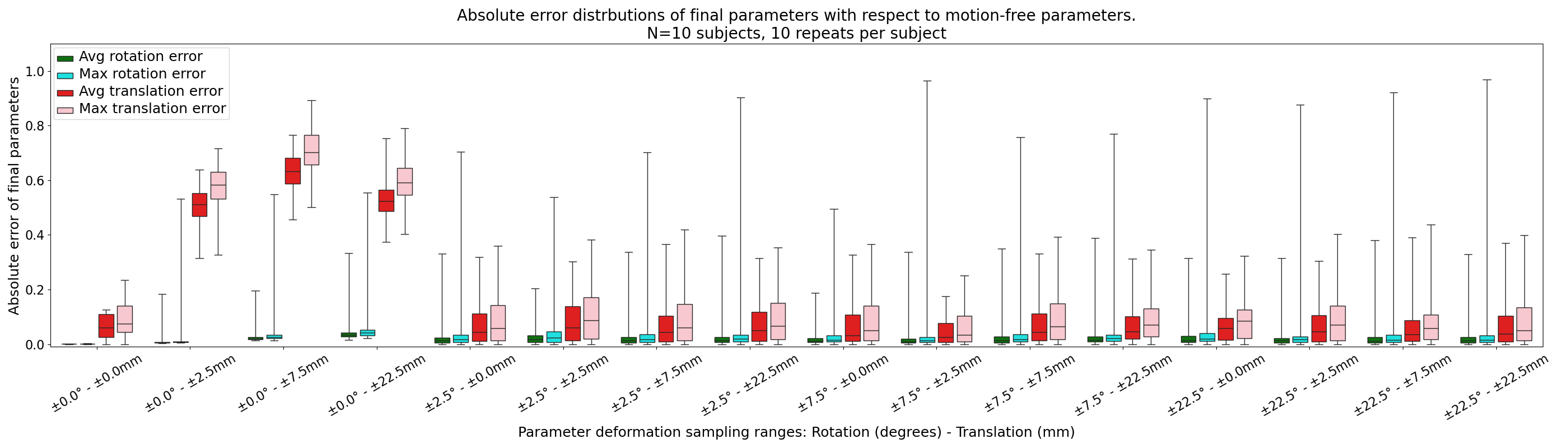}}
  \caption{Distributions of mean and maximum absolute error in parameters for a given optimization run (with respect to motion-free parameters). Absolute errors are presented both rotation and translation parameters. Results are shown under various ranges of uniformly sampled initial rigid transformations. Dataset consisted of 10 aligned subjects, each of which was randomly misaligned 10 times per condition.}
\label{fig:results}
\end{figure*}

We obtain sampling segment intensities $K^A$ and $K^B$ by projecting the sampling segment onto both image planes and bi-linearly interpolating pixel intensities at each coordinate. We do this for all $q$ time frames in the 2D+time slice, making $K^A$ and $K^B$ matrices of size $q\times p$. For each sampling line, we also create mask vectors  $\mathbf{m}^A, \mathbf{m}^B\in \{0, 1\}$ to denote whether a sample is within the bounds of a given image.

The loss is the sum of L2 intensity differences between pair-wise sampled lines across all point pairs and time frames:
\begin{equation}
    \mathcal{L}\left(\phi\right) = \sum_{A,B} \sum^q_j \sum^p_i \left(K^A_{ji} - K^B_{ji} \right)^2 * \left(\mathbf{m}^A_i * \mathbf{m}^B_i\right),
\end{equation}
where $\left(\mathbf{m}^A * \mathbf{m}^B\right)$ is used to exclude pixel-pairs with intensity samples outside image bounds.

The loss is minimized by computing the gradient w.r.t. the rotation and translation parameters $\phi$ through the affine transformation and intensity interpolation. Adam optimizer is used to increase convergence speed. We implement the algorithm in Pytorch and run it on an NVIDIA RTX2070 consumer-grade GPU card. We make the code repository publicly available\footnote{\href{https://github.com/NILOIDE/CMR_intensity_3Dmoco.git}{https://github.com/NILOIDE/CMR\_intensity\_3Dmoco.git}}.

\section{Experiments and results}
\label{sec:experiments}
As previously mentioned, we do not have any segmentation labels or contours available that could serve as a surrogate measure to evaluate the performance of the proposed method. Therefore, apart from demonstrating qualitatively that our method manages to recover the slice misalignments (Figure~\ref{fig:sample}) due to patient breathing/motion, we also investigate the convergence properties of the algorithm and present quantitative results on synthetic deformations.

\textbf{Dataset.} We evaluate the performance of our method using 10 CMR datasets from the UK Biobank\footnote{UK Biobank Imaging Study: \hyperref[http://imaging.ukbiobank.ac.uk]{http://imaging.ukbiobank.ac.uk}}. Particularly, each dataset is comprised of a stack of \(9-13\) individual SA slices with $1.8$\rm{mm} \(\times\) $1.8$\rm{mm} in-plane resolution and their corresponding \(2\)-chamber, \(3\)-chamber, and \(4\)-chamber LA scans with $1.8$\rm{mm} \(\times\) $1.8$\rm{mm} in-plane resolution. Additionally, each scan has \(50\) time frames. 

\textbf{Evaluation.} We use \(10\) motion-free datasets as our golden standard and we synthetically transform them using a random rigid transformation per slice and we recover the misalignment using the proposed method. In this manner, we conduct an algorithm stress-test by exercising control over the ranges. This enables us to assess the algorithm's robustness against both small and large misalignments.

We do this at various magnitudes of uniformly sampled perturbations of \(\pm 0.0^\circ, 2.5^\circ, 7.5^\circ, 22.5^\circ\) rotation and \(\pm0.0, 2.5, 7.5, 22.5\) \rm{mm} of translation around the motion-free parameters. We allow the algorithm 1000 steps to converge (approx. 30 seconds) before measuring its parameters' absolute error. In Figure~\ref{fig:results_before_after}, we report the distribution of absolute errors before and after optimization for each misalignment interval. Figure~\ref{fig:results} shows the post-optimization distributions of mean and max absolute errors of every run for the recovered rotation and translation parameters.

\textbf{Discussion.}
As seen in Figure~\ref{fig:results_before_after}, our algorithm reliably converges to the motion-free parameters even under the largest tested rotation and translation misalignments. Figure~\ref{fig:results} appears to show that our approach corrects rotations better than translations. Furthermore, the most challenging scenario is when only translations are present. We hypothesize that rotation motion is easier to correct as the intensities closer to the axes of rotation preserve enough intensity overlap to provide meaningful gradient information. This is however not the case with translation, where the entire intersection will stop displaying similarities in intensities given a large enough displacement, making gradient descent unable to discern an appropriate direction in which to shift the slices. Surprisingly, this phenomenon disappears when some rotation misalignment is also present.
\section{Conclusion}
This study explores the potential of intra-subject CMR slice alignment. For this reason, the proposed method involves aligning all SA and LA slices simultaneously. Unlike other approaches, our algorithm exclusively utilizes the image intensity information along slice intersections and requires no prior knowledge about the underlying anatomy. We demonstrate both quantitatively and qualitatively that the proposed method is capable of recovering a wide range of rigid transformations. 
However, our approach also demonstrates some limitations that we would like to address in the future. We believe it would be beneficial to utilize stochastic optimization techniques that would further allow the algorithm to overcome local minima. Moreover, we believe another beneficial addition would be to incorporate a loss function that is more robust against intensity changes across slices such as mutual information. 
\section{Acknowledgements}
This research study was conducted retrospectively using human subject data made available in open access by the UK Biobank Resource under Application Number 87802. Ethical approval was \textbf{not} required as confirmed by the license attached with the open access data.

This work is funded in part by the European Research Council (ERC) project Deep4MI (884622) and the Munich Center for Machine Learning.

\bibliographystyle{IEEEbib}
\bibliography{strings,refs}

\end{document}